\documentclass[conference]{IEEEtran}
\IEEEoverridecommandlockouts
\usepackage[nospace]{cite}
\usepackage{amsmath,amssymb,amsfonts}
\usepackage{graphicx}
\usepackage{textcomp}
\usepackage{xcolor}
\usepackage{subcaption}
\usepackage{bm}
\usepackage{url}
\usepackage{multirow}
\usepackage{pifont}
\usepackage{algorithm}
\usepackage{algpseudocode}
\usepackage{amssymb}
\usepackage{amsthm}
\usepackage[]{hyperref} 
\usepackage[flushleft]{threeparttable}

\def\BibTeX{{\rm B\kern-.05em{\sc i\kern-.025em b}\kern-.08em
    T\kern-.1667em\lower.7ex\hbox{E}\kern-.125emX}}

\usepackage{amsmath,amsfonts,bm}

\def\eqref#1{equation~\ref{#1}}

\def\1{\bm{1}}

\def\va{{\bm{a}}}

\def\vg{{\bm{g}}}

\def\vw{{\bm{w}}}

\DeclareMathAlphabet{\mathsfit}{\encodingdefault}{\sfdefault}{m}{sl}
\SetMathAlphabet{\mathsfit}{bold}{\encodingdefault}{\sfdefault}{bx}{n}

\def\sR{{\mathbb{R}}}

\def\sZ{{\mathbb{Z}}}

\DeclareMathOperator*{\argmin}{arg\,min}

\begin{document}

\title{Accelerating Distributed K-FAC with Smart Parallelism of Computing and Communication Tasks}

\author{\IEEEauthorblockN{Shaohuai Shi, Lin Zhang, Bo Li\\}
\IEEEauthorblockA{Department of Computer Science and Engineering, The Hong Kong University of Science and Technology\\
	shaohuais@cse.ust.hk, lzhangbv@connect.ust.hk, bli@cse.ust.hk}
}

\maketitle

\begin{abstract}

Distributed training with synchronous stochastic gradient descent (SGD) on GPU clusters has been widely used to accelerate the training process of deep models. However, SGD only utilizes the first-order gradient in model parameter updates, which may take days or weeks. Recent studies have successfully exploited approximate second-order information to speed up the training process, in which the Kronecker-Factored Approximate Curvature (KFAC) emerges as one of the most efficient approximation algorithms for training deep models. Yet, when leveraging GPU clusters to train models with distributed KFAC (D-KFAC), it incurs extensive computation as well as introduces extra communications during each iteration. In this work, we propose D-KFAC (SPD-KFAC) with smart parallelism of computing and communication tasks to reduce the iteration time. Specifically, 1) we first characterize the performance bottlenecks of D-KFAC, 2) we design and implement a pipelining mechanism for Kronecker factors computation and communication with dynamic tensor fusion, and 3) we develop a load balancing placement for inverting multiple matrices on GPU clusters. We conduct real-world experiments on a 64-GPU cluster with 100Gb/s InfiniBand interconnect. Experimental results show that our proposed SPD-KFAC training scheme can achieve 10\%-35\% improvement over state-of-the-art algorithms.
\end{abstract}

\begin{IEEEkeywords}
Distributed Deep Learning; K-FAC; Second-Order; Smart Parallelism; Load-Balancing
\end{IEEEkeywords}

\section{Introduction}\label{sec:intro}
Deep neural networks (DNNs) have been successfully deployed in numerous practical applications. However, due to the large size of deep models and the enormous amount of data involved, training a satisfactory model with the commonly used first-order algorithms such as stochastic gradient descent (SGD) and its variants is very time-consuming. To accelerate the training process of deep models, on one hand, distributed training techniques especially the distributed synchronized SGD (S-SGD) with data parallelism have become a common practice to train deep models with multiple processors~\cite{dean2012large,zhang2017poseidon,goyal2017accurate,you2018imagenet,jia2018highly,dong2020eflops,shi2021towards}; on the other hand, there exist many proposals to explore the second-order information like empirical Fisher Information Matrix (FIM) to accelerate the training process targeting at reducing the number of iterations~\cite{martens2015optimizing,grosse2016kronecker,botev2017practical,martens2018kronecker,george2018fast,osawa2019large,thomas2020interplay,goldfarb2020practical,lee2020continual}. The Kronecker-Factored Approximate Curvature (KFAC) has been successfully used as an approximate FIM to precondition the gradient through layer-wise block-diagonalization and Kronecker factorization for training large-scale convolutional neural networks (CNNs)~\cite{martens2015optimizing,grosse2016kronecker,george2018fast}. Osawa et al.,~\cite{osawa2019large,osawa2020scalable} show that distributed KFAC (D-KFAC) can achieve the target accuracy of ResNet-50~\cite{he2016deep} model on the ImageNet~\cite{deng2009imagenet} data set in $1/3$ number of epochs than the standard training with SGD. Note, however, that D-KFAC requires extensive computations to calculate preconditioning matrices compared to the first-order gradients, and also introduces significant communication costs on GPU clusters~\cite{osawa2019large,ueno2020rich}. 

Formally, the first-order SGD uses the update rule to minimize the objective function $f:\sR^{d}\to \sR$
\begin{equation}\label{equ:firstorder}
    \vw^{(t+1)}=\vw^{(t)}-\alpha^{(t)} \nabla \mathcal{L}(\vw^{(t)},\xi^{(t)}),
\end{equation}
where $\vw^{(t)}\in \sR^d$ is the model parameter, $\xi^{(t)}$ is a mini-batch of randomly sampled data, $\nabla\mathcal{L}(\vw^{(t)},\xi^{(t)}) \in \sR^d$ is the first-order stochastic gradient, and $\alpha^{(t)}$ is the learning rate at iteration $t$. On the other hand, the second-order method is to precondition the gradient $\vg^{(t)}$ with the inverse of FIM, i.e.,
\begin{equation}\label{equ:secondorder}
    \vw^{(t+1)}=\vw^{(t)}-\alpha^{(t)} F^{-1}(\vw^{(t)}) \nabla\mathcal{L}(\vw^{(t)},\xi^{(t)}),
\end{equation}
where $F^{-1}(\cdot)$ is the inverse of FIM. However, the dimension of $\vw^{(t)}$ in deep models would be very large (e.g., up to billions), which makes the inverse of FIM be impractical. The KFAC is an efficient approximation of FIM using layer-wise block-diagonalization and Kronecker factorization such that layer-wise matrices are easier to be inverted. To be specific, KFAC approximates $F$ with a diagonal block matrix $\hat{F}$, where each block matrix is corresponding to one layer of the DNN, i.e.,
\begin{equation}\label{equ:kfac}
    \hat{F}=diag(\hat{F}_1,\hat{F}_2,...,\hat{F}_L),
\end{equation}
where $\hat{F}_l$ is an approximate FIM of layer $l$ of an $L$-layer deep model. Thus, the update formula becomes
\begin{equation}\label{equ:kfacupdate}
    \vw^{(t+1)}_{l}=\vw^{(t)}_{l}-\alpha^{(t)} \hat{F}^{-1}(\vw_l^{(t)})\nabla\mathcal{L}_l(\vw^{(t)},\xi^{(t)}),
\end{equation}
where $\vw^{(t)}_{l}$ and $\nabla\mathcal{L}_l(\vw^{(t)},\xi^{(t)})$ are the model parameter and the gradient of layer $l$, respectively. For ease of presentation, we use $\hat{F}_l$ to represent $\hat{F}(\vw_l^{(t)})$. Compared to SGD, KFAC requires to calculate $\hat{F}_l$ and its inverse to precondition the gradient, which results in extra extensive computational costs. 

The existing state-of-the-art distributed KFAC (MPD-KFAC)~\cite{tsuji2019performance,osawa2019large,osawa2020scalable,pauloski2020convolutional,ueno2020rich} makes use of the concept of model parallelism with multiple GPUs to calculate the inverses of different layers' approximate FIMs in parallel to reduce the computing time. However, besides the communication in aggregating Kronecker factors, MPD-KFAC further introduces significant communication overheads in collecting inverted matrices. There were a number of attempts~\cite{osawa2019large,pauloski2020convolutional,ueno2020rich} to alleviate the communication overhead, but they fail to capture the parallelism between computing and communication tasks, which results in low throughput in a distributed system. It has been shown that communication tasks and computing tasks can be scheduled in S-SGD so as to hide some communication overheads to improve the system throughput~\cite{zhang2017poseidon,shi2019mg,peng2019generic,bao2020preemptive,shi2021mg}. Yet, the computation and communication paradigm of D-KFAC is much different from S-SGD so that the existing scheduling algorithms cannot be readily applied~\cite{shi2019mg,shi2020communication}. In this work, we first present a systemic performance analysis of D-KFAC to identify its performance bottlenecks. With the observation of pipelining between the communication tasks and computing tasks in D-KFAC, we propose a smart parallel solution, SPD-KFAC, to reduce the communication overheads and thus improve the system throughput. In particular, we propose two novel optimization methods in SPD-KFAC: 1) design a pipeline technique for the Kronecker factors computations and communications with tensor fusion; and 2) design a load-balance scheme for computations of inverting the Kronecker factors on multiple GPUs while incorporating the communication costs. We implement SPD-KFAC using the popular deep learning (DL) framework PyTorch\footnote{https://www.pytorch.org} and Horovod\footnote{https://github.com/horovod/horovod} and conduct extensive experiments on a 64-GPU cluster for performance evaluation with four modern CNNs. The experimental results show that SPD-KFAC achieves 10\%-35\% improvement over state-of-the-art methods. 

The rest of the paper is organized as follows. We introduce some background and related work in distributed DL with both first-order and second-order training algorithms in Section~\ref{sec:background}. Then we present the empirical evaluation results in the existing second-order training algorithms to characterize performance bottlenecks in Section~\ref{sec:characterizing}. Our design to address the communication problem in D-KFAC is demonstrated in Section~\ref{sec:spdkfac}, followed with the system implementation in Section~\ref{sec:implementation}. We conduct detailed experimental studies to show the effectiveness of our solution in Section~\ref{sec:experiments}. Finally, we conclude the paper in Section~\ref{sec:conclusion}.

\begin{figure*}[!ht]
	\centering
	\begin{subfigure}{\textwidth}
		\includegraphics[width=0.98\linewidth]{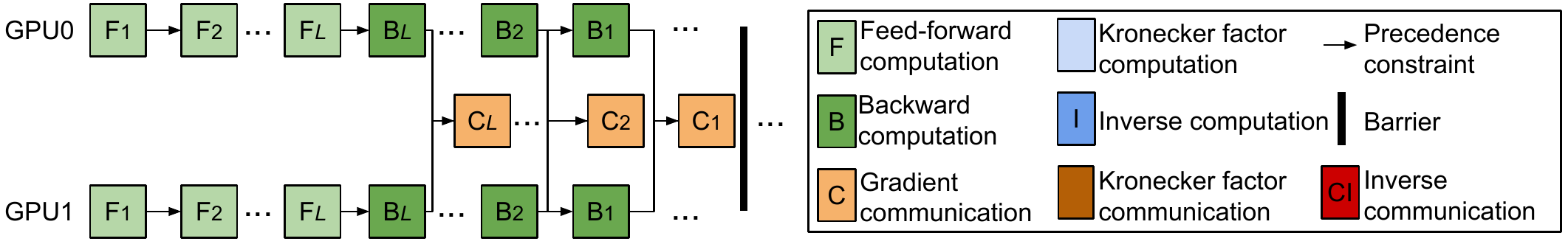}
		\caption{S-SGD: backward computations begin after all feed-forward computations have finished, while communication tasks of aggregating gradients can be paralleled with the backward computing tasks.}
	\end{subfigure}
	\begin{subfigure}{\textwidth}
	   \vspace{10pt}
		\includegraphics[width=0.98\linewidth]{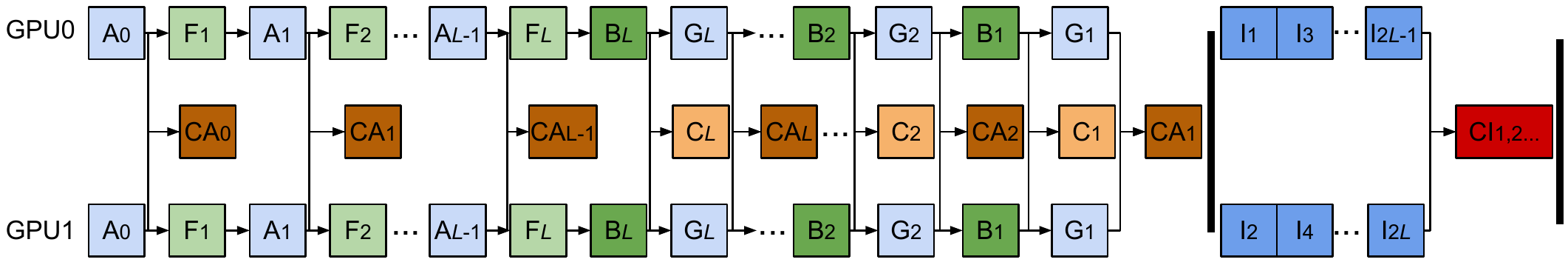}
		\caption{D-KFAC: the Kronecker factor computing task of $A$ (or $G$) is executed before (or after) the feed-forward (or backward) computation at that layer, and the communications of aggregating Kronecker factors can be paralleled with the computations. The computations of inverting matrices are placed on different GPUs, and the results are finally collected by all GPUs (i.e., MPD-KFAC).}
	\end{subfigure}
	\caption{The computation tasks (green and blue blocks) and communication tasks (brown and red blocks) in one iteration of S-SGD and D-KFAC on two GPUs.}
	\label{fig:preliminaries}
\end{figure*}
\section{Background and Related Work}\label{sec:background}

\subsection{Distributed SGD with Data Parallelism}
Distributed synchronized SGD (S-SGD) is one of the main training algorithms for large-scale DL due to its good convergence property and ease to scale-out~\cite{goyal2017accurate,you2018imagenet,jia2018highly,you2020large}. Compared to SGD in a single device, the update formula of S-SGD on a $P$-worker cluster becomes
\begin{equation}
    \vw^{(t+1)}=\vw^{(t)}-\frac{\alpha^{(t)}}{P} \sum_{p=1}^P\nabla \mathcal{L}(\vw^{(t)},\xi^{(t),p}),
\end{equation}
where $\nabla \mathcal{L}(\vw^{(t)},\xi^{(t),p})$ is the stochastic gradient at worker $p$ at iteration $t$ with locally sampled data $\xi^{(t),p}$. The aggregation of distributed gradients requires data communication between multiple workers (e.g., multiple GPUs in a single node and multiple GPUs in a distributed cluster), which is generally implemented with an all-reduce operation~\cite{awan2017s,goyal2017accurate,jia2018highly} or through parameter servers (PS)~\cite{li2014scaling,ba2017distributed}. The PS architecture requires central servers to collect gradients (and send model parameters) from (and to) multiple workers so that the central servers easily become the bandwidth bottleneck for large-scale training~\cite{yan2015performance,shi2018performance}. On the other hand, the all-reduce architecture, which has a long-time history from high performance computing community, has been widely used for large-scale training~\cite{goyal2017accurate,you2018imagenet,jia2018highly,you2020large,dong2020eflops}. In this work, we will also focus on the all-reduce architecture for data aggregation.

The main time-consuming parts of S-SGD on GPU clusters are feed-forward computation, backward propagation computation, and gradient aggregation as shown in Fig.~\ref{fig:preliminaries}(a). The first two parts are known as computing tasks, and the third part is known as the communication task.

To improve the system scalability, pipelining between computing tasks and communication tasks is one of the main methods to hide some communication overheads~\cite{awan2017s,zhang2017poseidon,shi2019mg,peng2019generic,bao2020preemptive}. Due to the layer-wise structure of deep models, the gradient aggregation of the current layer has no dependency with its previous layer's gradient calculation as shown in Fig.~\ref{fig:preliminaries}(a). Therefore, one can pipeline the communication tasks of gradient aggregation with the gradient computations during the backward pass, which is known as the wait-free backpropagation (WFBP) algorithm~\cite{awan2017s,zhang2017poseidon}. For example, the gradient communication of layer $l$ can be parallelized with the gradient computation of layer $l-1$. To alleviate the negative impact of the startup time of all-reduce operations, tensor fusion (or merged-gradient~\cite{shi2019mg,shi2021mg}) is necessary to merge nearby layers to be communicated together, which has been a default feature in distributed training frameworks like Horovod and TensorFlow.

\subsection{Distributed KFAC (D-KFAC)}
KFAC approximates the FIM as Kronecker products of small matrices which can be efficiently inverted as shown in Eq. (\ref{equ:kfac})~\cite{martens2015optimizing}. For each $\hat{F}_l$ at layer $l$ ($l=1,2,...,L$), it is calculated by
\begin{equation}
    \hat{F}_l=\va_{l-1}\va_{l-1}^\top \otimes \vg_l\vg_l^\top,
\end{equation}
where $\va_{l-1}$ and $\vg_l$ are the activation output at layer $l-1$ (or the input of layer $l$) and the gradient w.r.t. the output of layer $l$, respectively. $\otimes$ is the Kronecker product. Let 
\begin{equation}\label{equ:factorA}
    A_{l-1}=\va_{l-1}\va_{l-1}^\top
\end{equation}
and 
\begin{equation}\label{equ:factorG}
    G_{l}=\vg_{l}\vg_{l}^\top,
\end{equation}
which are called Kronecker factors and are symmetric. Then we have
\begin{equation}
    \hat{F}_l=A_{l-1} \otimes G_{l}.
\end{equation}
According to the property of the Kronecker product, the inverse of $\hat{F}_l$ can be represented by
\begin{equation}
    \hat{F}_l^{-1}=A_{l-1}^{-1} \otimes G_{l}^{-1}.
\end{equation}
Thus, Eq. (\ref{equ:kfacupdate}) becomes
\begin{equation}
    \vw^{(t+1)}_{l}=\vw^{(t)}_{l}-\alpha^{(t)} A_{l-1}^{-1}\otimes G_{l}^{-1}  \nabla\mathcal{L}_l(\vw^{(t)},\xi_t) .
\end{equation}
In practice, a Tikhonov regularization is generally required before inverting to achieve better convergence~\cite{grosse2016kronecker,osawa2019large,pauloski2020convolutional}. Therefore, the update rule of KFAC becomes
\begin{equation}
    \vw^{(t+1)}_{l}=\vw^{(t)}_{l}-\alpha^{(t)} (A_{l-1}+\gamma I)^{-1} \otimes (G_{l}+\gamma I)^{-1} \nabla\mathcal{L}_l(\vw^{(t)},\xi_t),
\end{equation}
where $\gamma$ is a damping parameter and $I$ is an identity.

When exploiting multiple GPUs to collaborate on training a single model with D-KFAC, the Kronecker factors of $A_{l-1}^p$ and $G_l^p$ at worker $p$ are different from other workers as the locally sampled data is different, so $A_{l-1}^p$ and $G_l^p$ should be aggregated before inverted. That is
\begin{align}\label{equ:kfacfinal}
     \vw^{(t+1)}_{l}=\vw^{(t)}_{l}- &\alpha^{(t)} (\frac{1}{P}\sum_{p=1}^PA_{l-1}^p+\gamma I)^{-1} \otimes \notag\\
    (&\frac{1}{P}\sum_{p=1}^PG_{l}^p+\gamma I)^{-1} \frac{1}{P}\sum_{p=1}^{P}\nabla\mathcal{L}_l(\vw^{(t)},\xi_t^p)  .
\end{align}
Compared to S-SGD, D-KFAC requires six extra time-consuming operations at each layer: four computing operations (compute Kronecker factors $A_{l-1}^p$ and $G_l^p$ and their inverses) and two communication operations (aggregation of $A_{l-1}^p$ and $G_l^p$). 

Due to the high computational cost of inverting matrices, recent work has proposed the distributed algorithm to reduce the computation time of inverting matrices~\cite{osawa2019large,pauloski2020convolutional,ueno2020rich}. As shown in Eq. (\ref{equ:kfacfinal}), the inverse operations of Kronecker factors at different layers have no dependency with each other. In existing state-of-the-art solutions~\cite{osawa2019large,pauloski2020convolutional,ueno2020rich}, the workloads of different layers in computing inverses are distributed to multiple GPUs (with a concept of model parallelism), and their results are finally gathered to all GPUs for preconditioning gradients. An example is shown in the right hand side of Fig.~\ref{fig:preliminaries}(b). GPU0 computes layers 1, 3, 5, ..., while GPU1 computes layers 2, 4, 6, ..., and finally the results are gathered. We refer to this method as MPD-KFAC. However, there is one critical problem in MPD-KFAC. As the inverses of different layers' Kronecker factors are computed on different GPUs, every GPU should broadcast its own results to all the other GPUs, which introduces significant communication overheads. When the communication overhead is larger than the benefit of distributed workload computing, MPD-KFAC would not bring any performance improvement. In this work, we will propose a load-balancing placement based on the computing and communication time to determine which layers' inverses should be distributed to multiple GPUs.

\section{Characterizing Performance Bottlenecks}\label{sec:characterizing}
In this section, we characterize the performance of KFAC on a single GPU and its distributed versions (D-KFAC and MPD-KFAC) using the ResNet-50 model on a 64-GPU cluster (16 nodes with 4 Nvidia RTX2080Ti GPUs per node) connected with 100Gb/s InfiniBand. The more details of the cluster configuration can be seen in Section~\ref{sec:experiments}. We use the metric of average iteration time in running 1,000 iterations with a mini-batch size of 32. 

\begin{figure}[!ht]
	\centering
	\includegraphics[width=\linewidth]{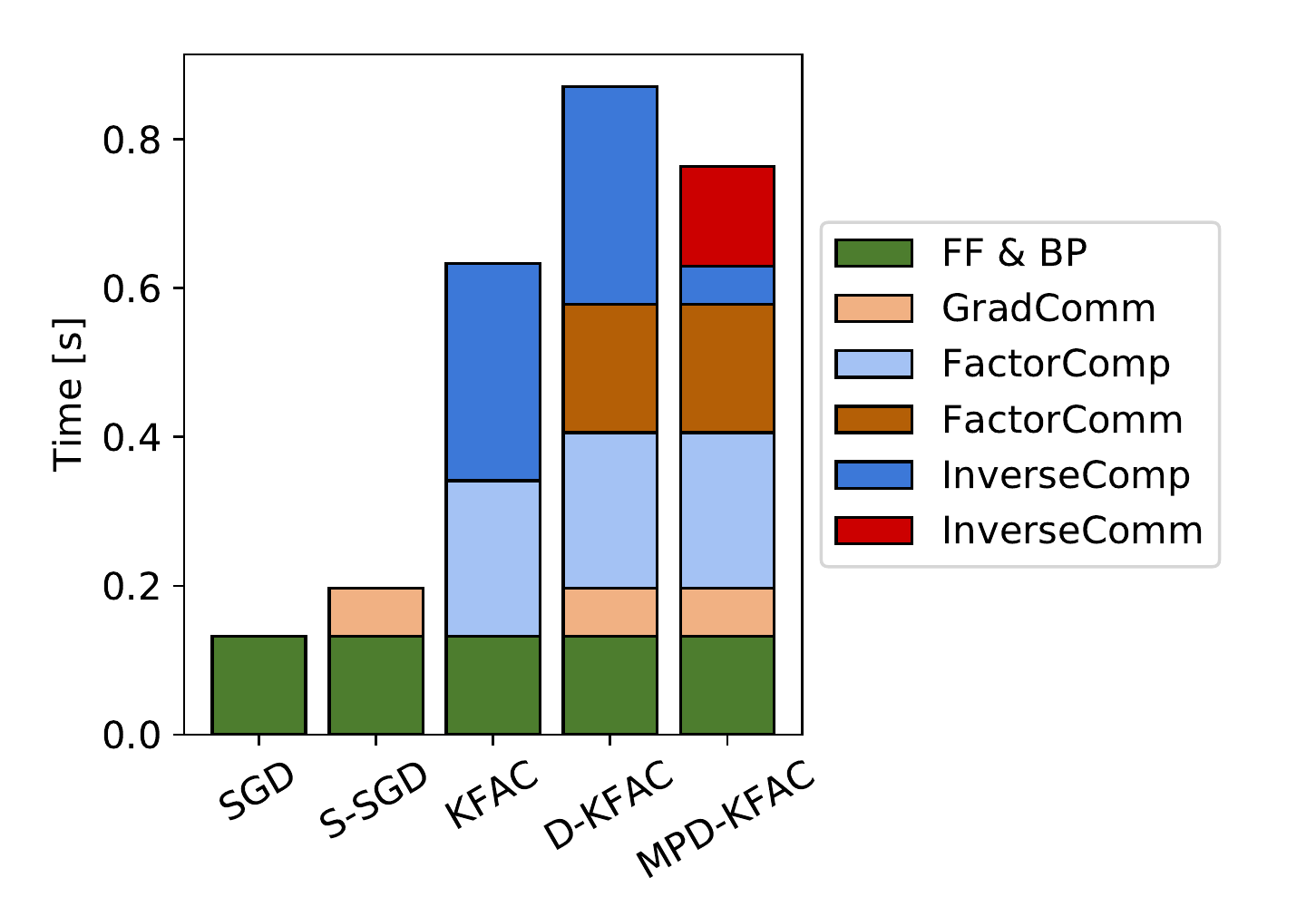}
	\caption{Time breakdowns of one iteration with existing training schemes. FF\&BP indicates the feed-forward and backpropagation computations. GradComm indicates the communication time of gradient aggregation. FactorComp and FactorComm indicate the computation and communication of factors ($A_{l-1}$ and $G_{l}$) respectively. InverseComp and InverseComm indicate the computation and communication of the inverse matrices respectively.}
	\label{fig:naivebreakdowns}
\end{figure}

The iteration time breakdowns of SGD and KFAC on a single GPU and their distributed versions (S-SGD, D-KFAC, and MPD-KFAC) on a 64-GPU cluster are shown in Fig.~\ref{fig:naivebreakdowns}. Note that the single-GPU algorithms do not introduce any communication overheads. It is seen that KFAC takes about 4 times slower than SGD due to the high computation workloads of constructing Kronecker factors and computing their inverses. When using multiple GPUs with D-KFAC, the communication cost of aggregating the distributed Kronecker factors (the dark brown block) is also much higher than aggregating gradients (the light brown block). The MPD-KFAC method, which distributes the inverse computations to multiple GPUs~\cite{osawa2019large,pauloski2020convolutional,ueno2020rich}, significantly reduces the computation time from 292ms to 51ms. However, the results of factors' inverses should be gathered by every GPU, which introduces dramatic communication overheads (around 134ms). In summary, there exist two main communication problems in the existing MPD-KFAC method: 1) high communication overheads of aggregating the Kronecker factors of $A_{l-1}$ and $G_{l}$, and 2) high communication overheads of broadcasting inverse matrices. 

\subsection{Communication of Kronecker Factors}
From the bar of MPD-FAC in Fig.~\ref{fig:naivebreakdowns}, we can see that the communication overhead of aggregating Kronecker factors $A_{l-1}$ and $G_{l}$ for $l=1,2,...,L$ is much higher than aggregating the gradients. The main reason is that the dimensions of $A_{l-1}$ and $G_{l}$, which are determined by the dimensions of activation outputs instead of model parameters~\cite{martens2015optimizing,grosse2016kronecker}, could be much higher than the number of parameters. In the evaluated ResNet-50 model, the number of model parameters is about 25.5 million, while the total elements of all $A_{l-1}$ and $G_{l}$ matrices (only the upper triangles are counted as the matrices are symmetric) are about 62.3 million and 14.6 million, respectively. Therefore, the communication traffic of Kronecker factors is much higher than gradients so that it requires a longer time for data aggregation. In this work, we will propose the pipelining technique to hide the communication costs of aggregating Kronecker factors.

\subsection{Communication of Matrices Inverses}
Using model parallelism to invert the Kronecker factors with multiple GPUs would introduce the communication overheads of sending inverses from a GPU to all other GPUs. Take the training of ResNet-50 on 64 GPUs as an example, we suppose that the first layer's inverses of Kronecker factors (i.e., $(A_0+\gamma I)^{-1}$ and $(G_1+\gamma I)^{-1}$) are computed by the first GPU. For ease of representation, we only use $A_{l-1}^{-1}$ and $G_{l}^{-1}$ to represent $(A_{l-1}+\gamma I)^{-1}$ and $(G_{l}+\gamma I)^{-1}$, respectively. Since the synchronized training algorithm requires that all GPUs should keep a consistent model at every iteration to guarantee the convergence, after $A_0^{-1}$ and $G_1^{-1}$ calculated, they should be sent to all other GPUs (can be finished by a broadcast operation) to precondition the gradients. So are the other layers' inverses. The dimensions of $A_{l-1}^{-1}$ and $G_{l}^{-1}$ are the same as $A_{l-1}$ and $G_{l}$, respectively, which are much larger than the dimension of gradients, so the communication traffic of broadcasting the results of inverses would be significant. 

In the bar of D-KFAC in Fig.~\ref{fig:naivebreakdowns}, there is no overhead of communicating the inverses of Kronecker factors as every GPU computes the inverses locally. Comparing MPD-KFAC with D-KFAC in terms of inverse computations, we can see that one can either distribute the workloads of inverse computations to multiple GPUs with some communication overheads or compute all inverses locally without any communication overheads but it requires high computation overheads. The computation time of inverting a matrix would be affected by the dimension of the matrix and the computing power of the hardware, while the communication time of broadcasting a matrix would be affected by the number of GPUs of the cluster and the bandwidth/latency of networks. It means that directly distributing all computation workloads of inverting matrices may not be optimal to reduce the overall time. We will propose a load-balancing placement approach to minimize the overhead of obtaining all the inverses of Kronecker factors.

\section{SPD-KFAC: Smart Parallel D-KFAC}\label{sec:spdkfac}
In this section, we present our proposed smart parallel D-KFAC (SPD-KFAC) to improve the throughput of distributed training with two novel optimizations: 1) alleviate the communication overhead of aggregating Kronecker factors by smartly pipelining the computing tasks and communication tasks, and 2) minimize the overhead of inverting Kronecker factors by balancing distributed workloads with considering the communication cost.

\subsection{Pipelining Computation and Communication of Kronecker Factors}\label{subsec:pipelining}
By exploiting the layer-wise structure of DNNs, we propose to pipeline the computing tasks of constructing the Kronecker factors and the communication tasks of aggregating the distributed Kronecker factors with tensor fusion. 

On one hand, the Kronecker factors $A_{l-1}^p$ ($l=1,2,...,L$) are computed according to Eq. (\ref{equ:factorA}) from the first layer to the last layer during the feed-forward pass. Once $A_{l-1}^p$ has been calculated, it can be aggregated with other workers directly so that the communication of $A_{l-1}^p$ can be overlapped with the computation of $A_{l}^p$. Ideally, if the communication cost of the Kronecker factor is comparable to the computation cost of its next layer's Kronecker factor, the communication of $A_{l-1}^p$ can be hidden by the computation of $A_{l}^p$. On the other hand, the Kronecker factors $G_{l}^p$ ($l=L,L-1,...,1$) are computed according to Eq. (\ref{equ:factorG}) from the last layer to the first layer during the backpropagation pass. Analogously to the pipelining of the forward pass, once $G_{l}^p$ has been calculated, it can be aggregated with other workers directly, so that its communication can also be overlapped with the computation of $G_{l-1}^p$.

\begin{figure}[!h]
	\centering
	\includegraphics[width=0.8\linewidth]{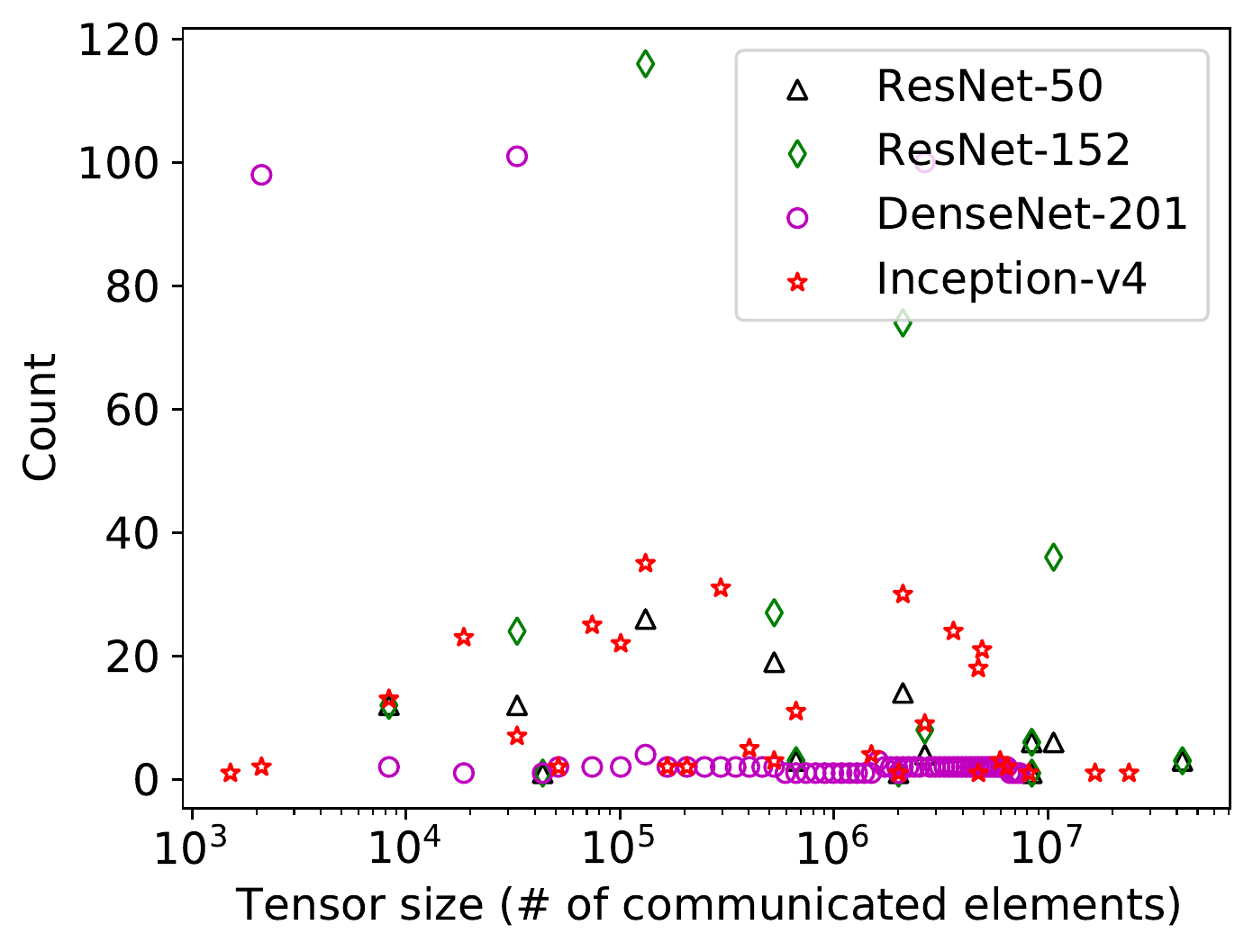}
	\caption{Tensor size distribution. The scatter markers indicate the number of tensors that have a specific size (the up triangle of symmetric Kronecker factors) in a DNN.}
	\label{fig:tensordistribution}
\end{figure}

Note that the dimensions of Kronecker factors in different layers are diverse. We plot the dimension distributions of Kronecker factors from four popular models in Fig.~\ref{fig:tensordistribution}. It is seen that, for example, in ResNet-50, the smallest number of communicated elements of the Kronecker factor is $2,080$ while the largest is $10,619,136$. Small tensors are difficult to fully utilize the bandwidth of networks due to the startup time of all-reduce operations~\cite{shi2019mg,jia2018highly}. We further propose dynamic tensor fusion for Kronecker factors. The tensor fusion technique applied in gradients is successfully used in large-scale training to improve the system scalability~\cite{shi2019mg,jia2018highly}. In particular, Shi et al.,~\cite{shi2019mg} proposed an optimal tensor fusion solution according to the gradient computation time and communication estimation time. Similarly, we exploit the layer-wise factor computation time, which can be measured through several iterations' running, and the dimension of factors to estimate the communication time to determine how to merge consecutive factors (e.g., $A_{l-1}^p$ and $A_{l}^p$) to be communicated once so that they can be better hidden by the simultaneous computing tasks. Different from MG-WFBP~\cite{shi2019mg}, the tensor fusion is applied on the Kronecker factors (i.e., $A_{l-1}^p$ and $G_{l}^p$) but not gradients, and it should be applied in both the feed-forward pass (for $A_{l-1}^p$) and the backward pass (for $G_{l}^p$).

\begin{figure}[!ht]
	\centering
	\includegraphics[width=\linewidth]{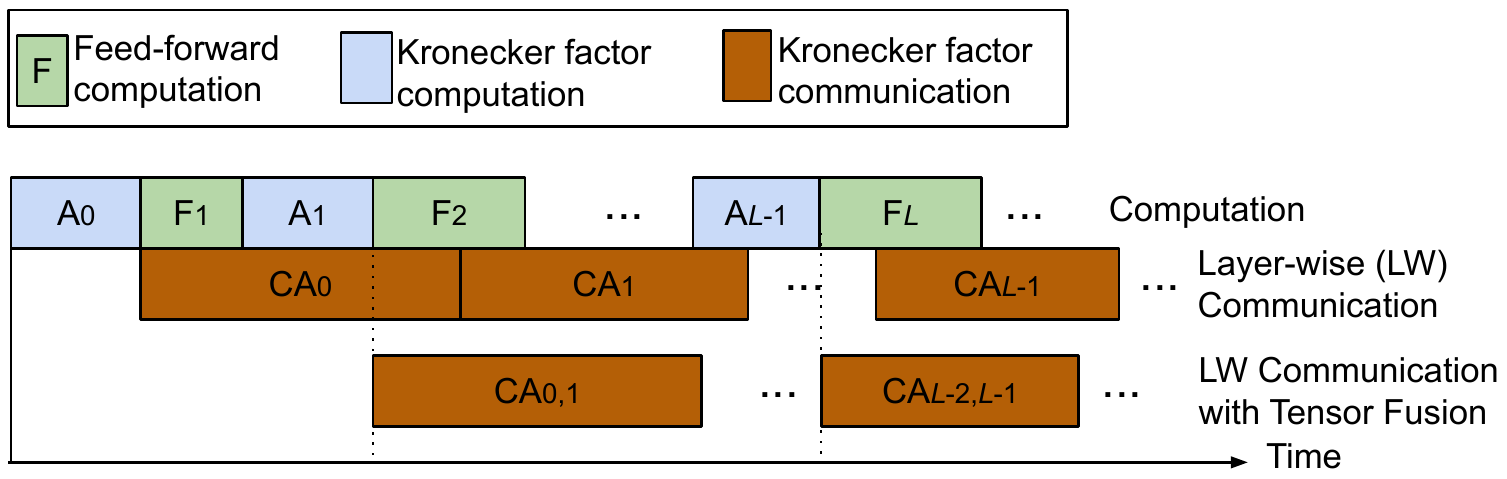}
	\caption{The pipeline between computation and communication of Kronecker factors. Using tensor fusion, some Kronecker factors are merged to be communicated once, e.g., $A_0$ and $A_1$ are communicated together.}
	\label{fig:tensorfusion}
\end{figure}

\textbf{Determining tensor fusion of Kronecker factors.} Formally, let $t_{Ap}^{(l)}$ and $t_{Am}^{(l)}$ denote the com\underline{p}utation time of calculating the Kronecker factor $A_{l}$ and its com\underline{m}unication time, respectively. The beginning times of computation and communication of $A_{l}$ are denoted by $\tau_{Ap}^{(l)}$ and $\tau_{Am}^{(l)}$, respectively. We use $t_f^{(l)}$ to denote the feed-forward computation time of layer $l$. We also assume that the communication time of an all-reduce operation follows the form
\begin{equation}\label{equ:allreduce-model}
    t_c(m) = \alpha_{ar}+\beta_{ar}\times m,
\end{equation}
where $m$ is the number of elements of the all-reduced vector, and $\alpha_{ar}$ and $\beta_{ar}$ are two environment-related parameters~\cite{shi2019mg,jia2018highly}. During the communication of the symmetric Kronecker factor $A_{l}\in \sR^{d_l\times d_l}$, whose number of communicated elements is $d_l\times(d_l+1)/2$, the corresponding computing tasks are the feed-forward computation of layer $l$ and the factor $A_{l}$ computation of layer $l+1$. Thus, according to the optimal tensor fusion in~\cite{shi2019mg}, if 
\begin{equation}\label{equ:fusioncriteria}
    \tau_{f}^{(l+1)}+t_f^{(l+1)}+t_{Ap}^{(l+1)} < \tau_{Am}^{(l)}+\alpha_{ar},
\end{equation}
where $\tau_{f}^{(l+1)}$ is the beginning time of feed-forward computation of layer $l$, we prefer to merge $A_{l}$ with $A_{l+1}$ to be communicated once. An illustrated example is shown in Fig.~\ref{fig:tensorfusion}. During the backward pass to compute and communicate the Kronecker factors $G_{l}$, we use the same policy to generate the tensor fusion solution for the communications of $G_{l}$.

Note that our proposed pipelining solution is very different from the methods presented~\cite{pauloski2020convolutional,ueno2020rich}. In~\cite{pauloski2020convolutional}, the communications of aggregating of Kronecker factors happen after all layers Kronecker factors have been calculated, which means that there is no opportunity for pipelining the communication tasks with computing tasks. On the other hand, in~\cite{ueno2020rich}, it also exploited the pipelining technique to overlap communications and computations, but it used a Reduce-ScatterV operation to perform the aggregation of all Kronecker factors $A_{l-1}^p$, which means the communications can only begin after all $A_{l-1}^p$ have been calculated and the communications can only be pipelined with the backward pass.

\subsection{Distributed Inverting Kronecker Factors}\label{subsec:lbp}
After the aggregation of distributed local Kronecker factors $A_{l-1}^p$ and $G_{l}^p$, all GPUs have consistent global Kronecker factors $A_{l-1}$ and $G_{l}$ for $l=1,2,...,L$. Before preconditioning gradients, all global Kronecker factors should be inverted by adding a damped identity, i.e., computing $(A_{l-1}+\lambda I)^{-1}$ and $(G_{l}+\lambda I)^{-1}$. The challenge of distributed inverting these matrices is how to place the workloads to different GPUs such that the overall elapsed-time for collecting all results on all GPUs is minimal. We formulate this problem as an optimization problem, and then we propose an efficient solution without affecting the training speed.

\textbf{Problem formulation. } Let $T_i\in \sR^{d_i\times d_i}$ ($i=1,2,...,N$) be $N$ tensors located at $P$ GPUs. There is a function $f: \sR^{d\times d}\to \sR^{d\times d}$, which is a computing-intensive operation, should be applied to all $N$ tensors so that all GPUs can obtain the results of $f(T_i)$ ($i=1,2,...,N$). Thus, the placement of workloads can be generalized as: a set of tensors $S_p$ is placed on the $p^{th}$ GPU ($p=1,2,...,P$), where
\begin{equation}
    S_1 \cup S_2 \cup ... \cup S_P = \{T_1, T_2, ..., T_N\}.
\end{equation}
The tensor $T_i$ should be placed either on only one GPU or on all GPUs, i.e., if
\begin{equation}
    \exists i, j, k, T_i\in S_j \text{ and } T_i \in S_k,
\end{equation}
where $1\leq i \leq N$, $1\leq j, k \leq P$, and $j \neq k$, then
\begin{equation}\label{equ:noncommunicated}
    T_i\in S_p, p=1,2,...,P,
\end{equation}
otherwise
\begin{equation}
    \forall j\neq k, S_j \cap S_k = \varnothing.
\end{equation}
For any tensor $T_i$ that satisfies Eq. (\ref{equ:noncommunicated}), all GPUs need to compute $f(T_i)$, but $f(T_i)$ does not need to be communicated. We denote such tensors as \textit{non-communicated} tensors (NCTs). Otherwise, we denote the tensors that do not satisfy Eq. (\ref{equ:noncommunicated}) as \textit{communicated} tensors (CTs). Let $t_i^{comp}$ be the computation time consumed by calculating $f(T_i)$ on a GPU, and $t_i^{comm}$ be the communication time of broadcasting the tensor $f(T_i)$ from one GPU to all other GPUs. In general, $t_i^{comp}$ and $t_i^{comm}$ are proportional to the dimension $d_i\times d_i$, i.e.,
\begin{equation}
    t_i^{comp} \propto d_i^2 \text{ and }t_i^{comm} \propto d_i^2.
\end{equation}
Then the overall time of obtaining all results by all GPUs can be generalized as 
\begin{equation}
    t=\max\{\sum_{i_p} t_{i_p}^{comp} + \sum_{j_p} t_{j_p}^{comm}\},
\end{equation}
where $i_p\in \{i|T_i\in S_p\}$, $j_p\in \{j|T_j\in S_p \text{ and } T_j \text{ is CT}\}$, and $1\leq p \leq P$. Our target is to generate the tensor sets $S_p$ ($p=1,2,...,P$) for all GPUs such that $t$ is minimal.

\begin{figure}[!h]
	\centering
	\includegraphics[width=\linewidth]{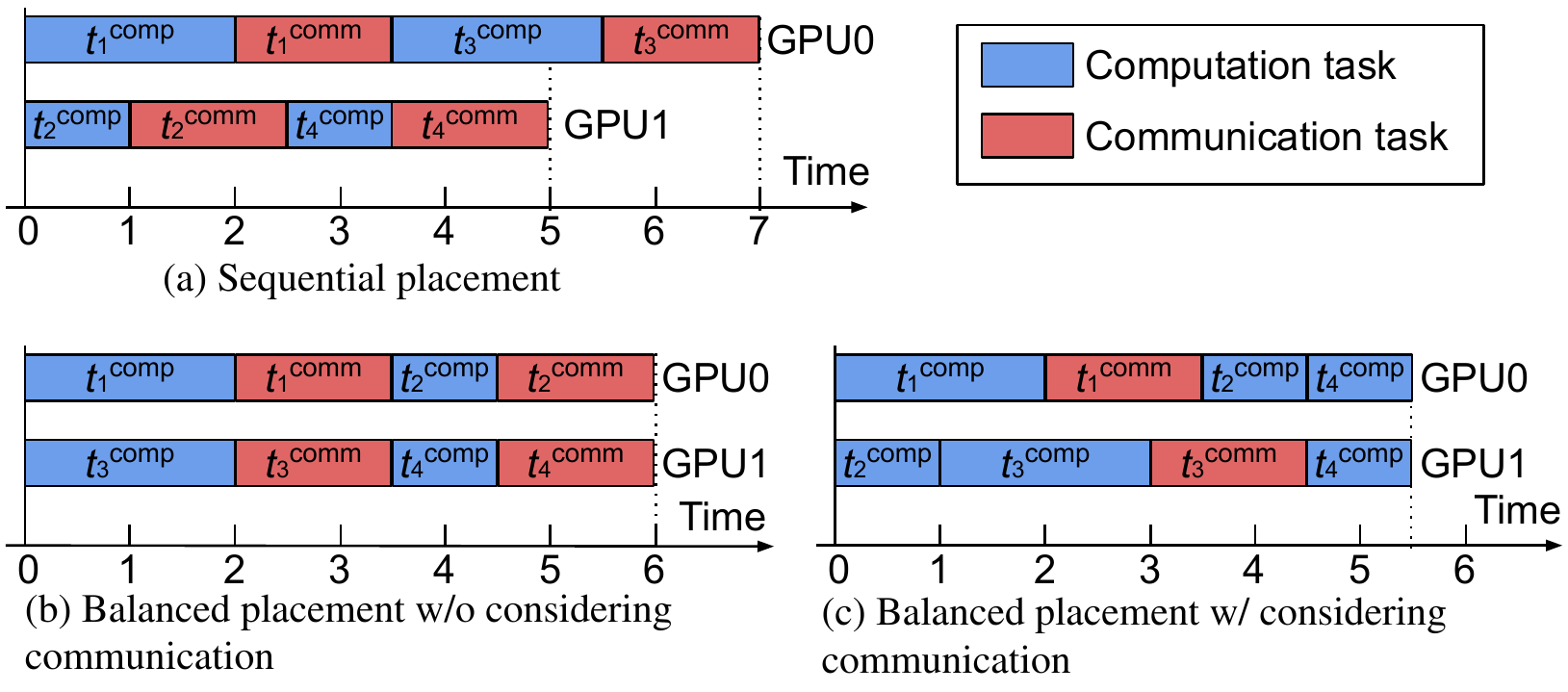}
	\caption{Examples of different placement solutions for four tensors on two GPUs.}
	\label{fig:placementexamples}
\end{figure}

In previous work~\cite{osawa2019large,pauloski2020convolutional,ueno2020rich}, the workloads of inverting $A_{l-1}$ and $G_{l}$ (totally $2L$ tensors, which are denoted as $T_1, T_2, ..., T_{2L}$) are placed sequentially to $P$ GPUs. That is, for the $p^{th}$ GPU, 
\begin{equation}
    S_p=\{T_{i}|i \% p=0 \text{ and } 1\leq i \leq 2L\},
\end{equation}
and all tensors are CTs, i.e.,
\begin{equation}
    \forall j\neq k, S_j \cap S_k = \varnothing.
\end{equation}
With such a placement solution, if $2L<P$, some GPUs would be idle without computing any tensors. As all tensors are CTs, all tensors' results should be communicated to other GPUs, thus the overall time is
\begin{equation}
    t=\max\{\sum_{i_p} t_{i_p}^{comp} + \sum_{i_p} t_{i_p}^{comm}\},
\end{equation}
where $i_p\in \{i|T_i\in S_p\}$. An example of four tensors on two GPUs is illustrated in Fig.~\ref{fig:placementexamples}(a). As the tensors are placed sequentially, different GPUs may have largely variant workloads (i.e., unbalanced workloads) so that one GPU spends a long time in computation and communication while other GPUs are idle. 

\textbf{Load-balancing placement}. 
To address the load-balancing problem in distributing multiple tensors to multiple GPUs, we propose a simple yet efficient algorithm to place the workload according to the dimensions of tensors. Since $t_i^{comp} \propto d_i^2$ and $t_i^{comm} \propto d_i^2$, which will be empirically verified in Section~\ref{sec:experiments}, we distribute the workloads to all GPUs in an uniform way according to $d_i^2$ ($i=1,2,...,N$). Formally, we would like to generate a solution of $S_p$ such that
\begin{equation}
    \sum_{i_1}d_{i_1}^2 \approx \sum_{i_2}d_{i_2}^2 \approx ... \approx \sum_{i_P}d_{i_P}^2,
\end{equation}
where $i_p \in \{i|T_i \in S_p\}$ and $p=1,2,...,P$. An example of the load-balancing placement is illustrated in Fig.~\ref{fig:placementexamples}(b), which saves one time slot compared to the sequential placement.

\textbf{Determining a tensor to be CT or NCT}.
On the other hand, some tensors may take longer time in communications than in computations if the dimension of the tensor is too small to utilize the network bandwidth. Therefore, instead of making all tensors be CTs, we use a simple policy according to the computation and communication time estimations to determine which tensors should be NCTs. That is, $T_i$ should be NCT if $t_i^{comp} < t_i^{comm}$, otherwise $T_i$ should be CT. An example of balancing workload placement with considering the communication cost is illustrated in Fig.~\ref{fig:placementexamples}(c), which saves 0.5 time slot compared to that without considering the communication cost.

In summary, our algorithm of load-balancing placement with dynamically determining tensor types is shown in Algorithm~\ref{algo:balancedplacement}. We first sort the dimensions of input tensors (Line 3), and then traversing the sorted dimension to put the current largest dimension tensor (Line 5) to the GPU that has a minimal workload (Line 11-13) if $T_i$ is CT, otherwise put the current tensor to all GPUs according to the estimated computation and communication time (Line 8-10). Note that Algorithm~\ref{algo:balancedplacement} has a time complexity of $O(N)$, and it only needs to be executed once to generate the workload set for each GPU at the beginning of training.

\begin{algorithm}[!ht]
	\caption{LBP: Load-Balancing Placement with Dynamically Determining Tensor Types}\label{algo:balancedplacement}
	\textbf{Input: } $[T_1,T_2...,T_n]$, $P$\\
	\textbf{Output: } $[S_1, S_2, ..., S_P]$
	\begin{algorithmic}[1]
		\small
		\State Initialize $S_p= \varnothing$ for $p=1,2,...,P$;
		\State Initialize $\bm{b}=[0] \in \sZ^P$; \Comment{A bucket array to record the workload has been assigned for a GPU.}
	    \State $\bm{I}=\text{argsort}([d_1,d_2,...,d_N])$;\Comment{Sort the dimensions in a descending order, and get their indices.}
	    \For{$i:= \bm{I}[1]\to \bm{I}[N]$}
	        \State $p=\argmin_p \bm{b}$;
	        \State $t_i^{comp}=\text{EstimateComputeTime}(d_i)$;
	        \State $t_i^{comm}=\text{EstimateCommunicationTime}(d_i)$;
	        \If{$t_i^{comp} <t_i^{comm}$}
	            \State Insert $T_i$ to $S_1, S_2, ..., S_P$;
	            \State $\bm{b}[1...P] = \bm{b}[1...P]+d_i$;
            \Else
	            \State Insert $T_i$ to $S_p$;
	            \State $\bm{b}[p] = \bm{b}[p]+d_i$;
	        \EndIf
	    \EndFor
	    \State Return [$S_1, S_2, ..., S_P$];
	\end{algorithmic}
	
\end{algorithm}

\section{Implementation}\label{sec:implementation}
\begin{figure}[!h]
	\centering
	\includegraphics[width=\linewidth]{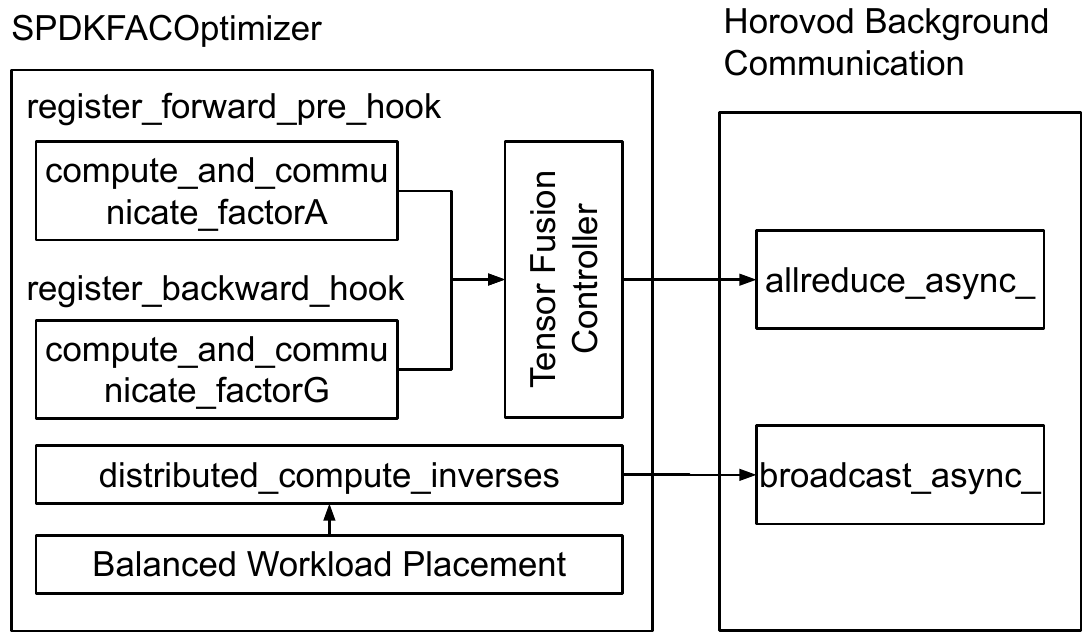}
	\caption{The overview of SPD-KFAC implementation.}
	\label{fig:implementaion}
\end{figure}

We have implemented our SPD-KFAC atop PyTorch and Horovod, where PyTorch is a general-purpose framework for DNN training and Horovod is used as an asynchronous communication framework to support pipelining between computing tasks and communication tasks. To keep the standard APIs of DNN training in PyTorch, we inherit the class ``torch.optim.Optimizer'' as a new class named ``SPDKFACOptimizer'' so that only one extra line of code should be inserted to support our proposed SPD-KFAC training algorithm. The overview of our SPD-KFAC implementation in ``SPDKFACOptimizer'' is shown in Fig.~\ref{fig:implementaion}.

\subsection{Pipelining Factors Computation and Communication}
To support the pipeline between Kronecker factors computations and communications, we exploit the ``register\_forward\_pre\_hook'' API in PyTorch to register the factor computation function (``compute\_and\_communicate\_factorA'') which would be executed before the feed-forward computation of the current layer. In ``compute\_and\_communicate\_factorA'', we first compute the factor $A_{l}$ in a synchronous way, and then send $A_{l}$ (using the object reference without copying data) to our ``TensorFusionController'' which determines whether the current coming tensor should be communicated directly or fused with other tensors. In ``TensorFusionController'', it determines whether the being communicated tensors should be fused tensors according to Eq. (\ref{equ:fusioncriteria}) and sends the ready tensors to the Horovod asynchronous all-reduce API (i.e., ``hvd.allreduce\_async\_''). Therefore, the asynchronous all-reduce communications can be parallelized with the computations of Kronecker factors. During the backward propagation, similarly to the hook of feed-forward, we exploit the ``register\_backward\_hook'' API to register the factor computation function (``compute\_and\_communicate\_factorG'') which would be executed after the backward computation of the current layer. In the hook function, the Kronecker factor $G_{l}$ of the current layer is computed then sent to ``TensorFusionController''.

\subsection{Load-Balancing Placement for Inverting Kronecker Factors}
To generate the load-balancing placement of inverting matrices on multiple GPUs, we first build a computation performance model (CompPM) and a communication performance model (CommPM) for computing a matrix inverse on a GPU and broadcasting a tensor on a GPU cluster, respectively. First, we measure the computation time of a serial of tensors $T\in \sR^{d\times d}$, where $d\in[64,8192]$, on a particular GPU. The chosen range can cover most matrix dimensions in real-world DNNs. Using the measured data, we fit them to an exponential function of
\begin{equation}\label{equ:inverting-model}
    t^{comp}(d) = \alpha_{inv}\times e^{\beta_{inv} \times d}.
\end{equation}
Second, we measure the communication time of broadcasting a tensor with dimension $d\in[64,8192]$ and fit the data with a function of
\begin{equation}\label{equ:broadcast-model}
    t^{comm}(d) = \alpha_{bcast}+\beta_{bcast} \times d(d+1)/2.
\end{equation}
For a particular GPU cluster, we only need to estimate $\alpha_{inv}$, $\beta_{inv}$, $\alpha_{bcast}$, and $\alpha_{bcast}$ for the above two equations through one-time benchmarking.

With the configured DNN (``nn.Module'') in PyTorch, the dimensions of Kronecker factors can be achieved by traversing all layers so that Algorithm~\ref{algo:balancedplacement} can be executed during the initialization of ``SPDKFACOptimizer''. During training, after $A_{l-1}$ and $G_{l}$ have been aggregated, all GPUs invoke the computations of inverting assigned tensors (i.e., $T_i\in S_p$). The inverse operation is natively supported by the highly optimized library cuSolver\footnote{\url{https://docs.nvidia.com/cuda/cusolver/index.html}} on Nvidia GPUs, which exploits the Cholesky decomposition to compute the inverse and it can be easily integrated to PyTorch using CuPy~\cite{cupy_learningsys2017}. After inverting a tensor on a GPU, the inverted result is directly broadcasted using the Horovod asynchronous broadcast API (``hvd.broadcast\_async\_'') to all other GPUs if the tensor is CT. When some GPUs are assigned with multiple tensors, it is also possible that the communication task of the current tensor is overlapped with the computing task of other tensors, which would bring some performance gain.

Note that the $A_{l-1}$ and $G_{l}$ are both symmetric, so their inverse matrices $A_{l-1}^{-1}$ and $G_{l}^{-1}$ should be also symmetric. When broadcasting $A_{l-1}^{-1}$ and $G_{l}^{-1}$, we only need to send their upper triangle elements (including the diagonals) so that the communication traffic can be reduced to $d(d+1)/2$ for a matrix $A^{-1} \text{ or } G^{-1} \in \sR^{d\times d}$. 

\section{Experimental Studies}\label{sec:experiments}
In this section, we first evaluate some parameters of Eq.~(\ref{equ:allreduce-model}), Eq.~(\ref{equ:inverting-model}), and Eq.~(\ref{equ:broadcast-model}) for the communication model of all-reduce, the computation model of inverting matrices, and the communication model of broadcast on our testbed. Then we show the experimental studies compared to existing solutions in~\cite{osawa2019large,pauloski2020convolutional,ueno2020rich}, and we also provide an ablation study of our proposed method. Since our proposed algorithms are systemic optimizations without affecting the numerical results of D-KFAC, our training algorithm SPD-KFAC should generate identical numerical results and thus has the same convergence properties as D-KFAC. Therefore, we will not compare the convergence since it has been well-verified in~\cite{osawa2019large,pauloski2020convolutional,ueno2020rich}. 

\subsection{Experimental Settings}
\textbf{Testbed.} Our testbed is a 64-GPU cluster of 16 nodes, in which each node is equipped with 4 Nvidia RTX2080Ti GPUs connected PCIe3.0x16. The interconnect between nodes is 100Gb/s InfiniBand which supports RDMA transports for all-reduce and broadcast operations. The configuration details of each node are shown in Table~\ref{table:nodeconfig}. The common software related to computation and communication performance are Nvidia GPU Driver-440.36, CUDA-10.2,  OpenMPI-4.0.1\footnote{\url{https://www.open-mpi.org/}}, NCCL-2.4.7\footnote{\url{https://developer.nvidia.com/nccl}}, Horovod-1.9.2, and PyTorch-1.4.0 with cuDNN-7.6. 
\begin{table}[!ht]
	\centering
		\caption{The server configuration.}
		\label{table:nodeconfig}
		\begin{tabular}{|l|l|}
			\hline
			Name & Model \\\hline\hline
			CPU & Dual Intel(R) Xeon(R) Gold 6230 CPU @ 2.10GHz \\
			GPU & Nvidia RTX2080Ti (@1.35GHz and 11GB Memory)\\
			Memory & 512GB DDR4 \\
			Network & 100Gb/s InfiniBand \\
			OS & Ubuntu-16.04\\\hline
		\end{tabular}
\end{table}

\textbf{DNN models.} Similar to~\cite{osawa2019large,pauloski2020convolutional,ueno2020rich}, we choose the ResNet architectures~\cite{he2016deep} including ResNet-50 and ResNet-152 with the ImageNet data set~\cite{deng2009imagenet} to evaluate the training performance. In addition, we further select two other CNNs DenseNet-201~\cite{huang2017densely} and Inception-v4~\cite{szegedy2017inception}. The details of the chosen models are shown in Table~\ref{table:dnnconfig}, where the batch size indicates the number of samples used on each GPU at each iteration and it is set for maximally utilizing the GPU memory. The input size of the input image resolution is $3\times 224\times 224$.

\begin{table}[!ht]
	\centering
		\caption{DNN details for experiments. ``\# Layers'' represents the number of layers that should be preconditioned in KFAC. ``\# As'' and ``\# Gs'' represent the total number of up triangle elements of Kronecker factors $A$ and $G$ respectively.}
		\label{table:dnnconfig}
    \begin{tabular}{|c|c|c|c|c|c|}
    	\hline
        \multirow{2}{*}{Model} & \# Param. & \multirow{2}{*}{\# Layers} & Batch & \# As & \# Gs \\
        & (million) & & Size & (million) & (million) \\\hline\hline
        ResNet-50& 25.6 & 54 & 32 & 62.3 & 14.6 \\\hline
    	ResNet-152 & 60.2 & 156 & 8 & 162.0 & 32.9 \\\hline
    	DenseNet-201& 20.0 & 201 & 16 & 131.0 & 18.0 \\\hline
    	Inception-v4& 42.7 & 150 & 16 & 116.4 & 4.7 \\\hline
    \end{tabular}
\end{table} 

\subsection{Performance Models}
\begin{figure}[!ht]
	\centering
	\begin{subfigure}{0.24\textwidth}
		\includegraphics[width=\linewidth]{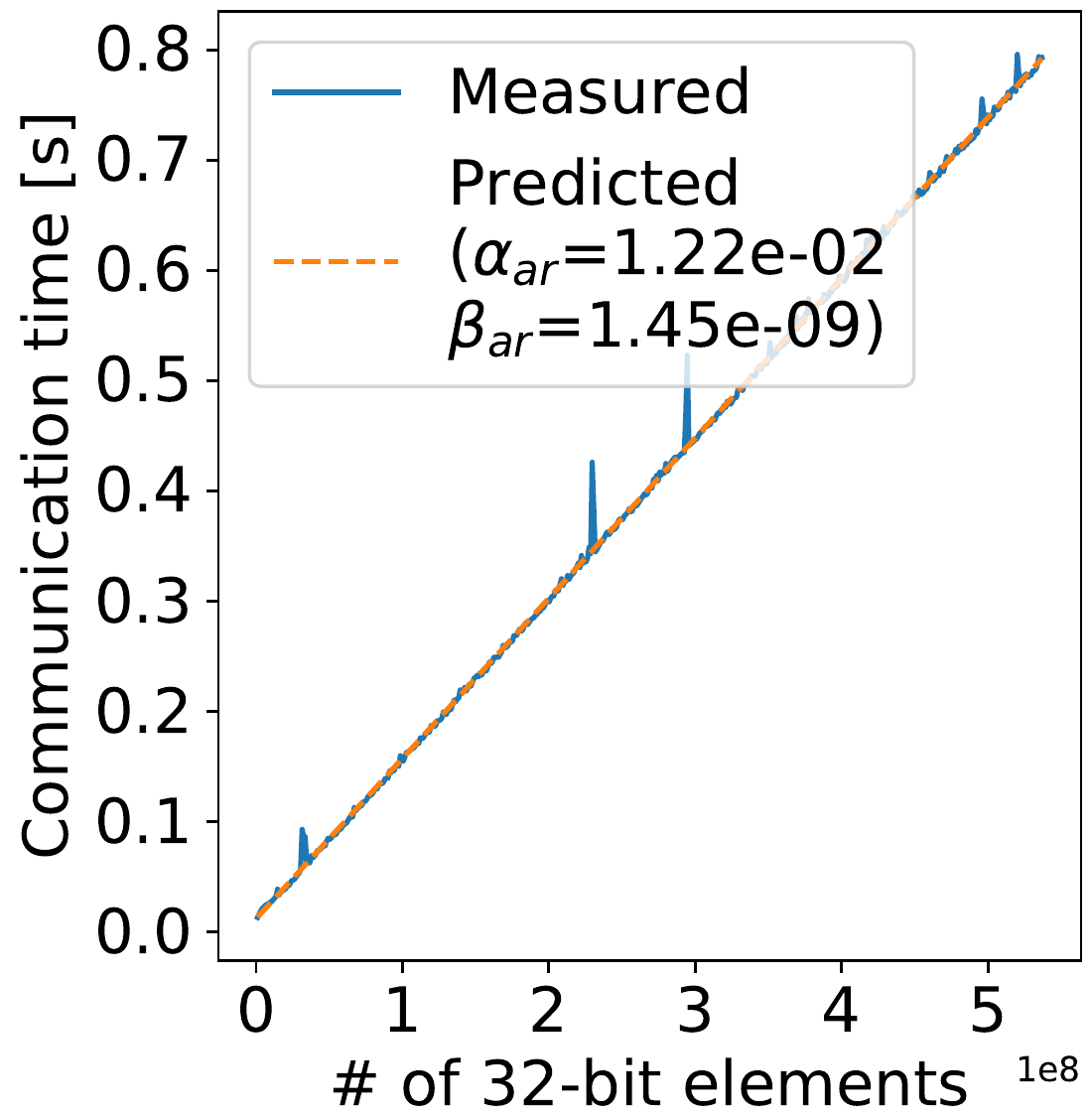}
		\caption{All-reduce.}
	\end{subfigure}
	\begin{subfigure}{0.24\textwidth}
		\includegraphics[width=\linewidth]{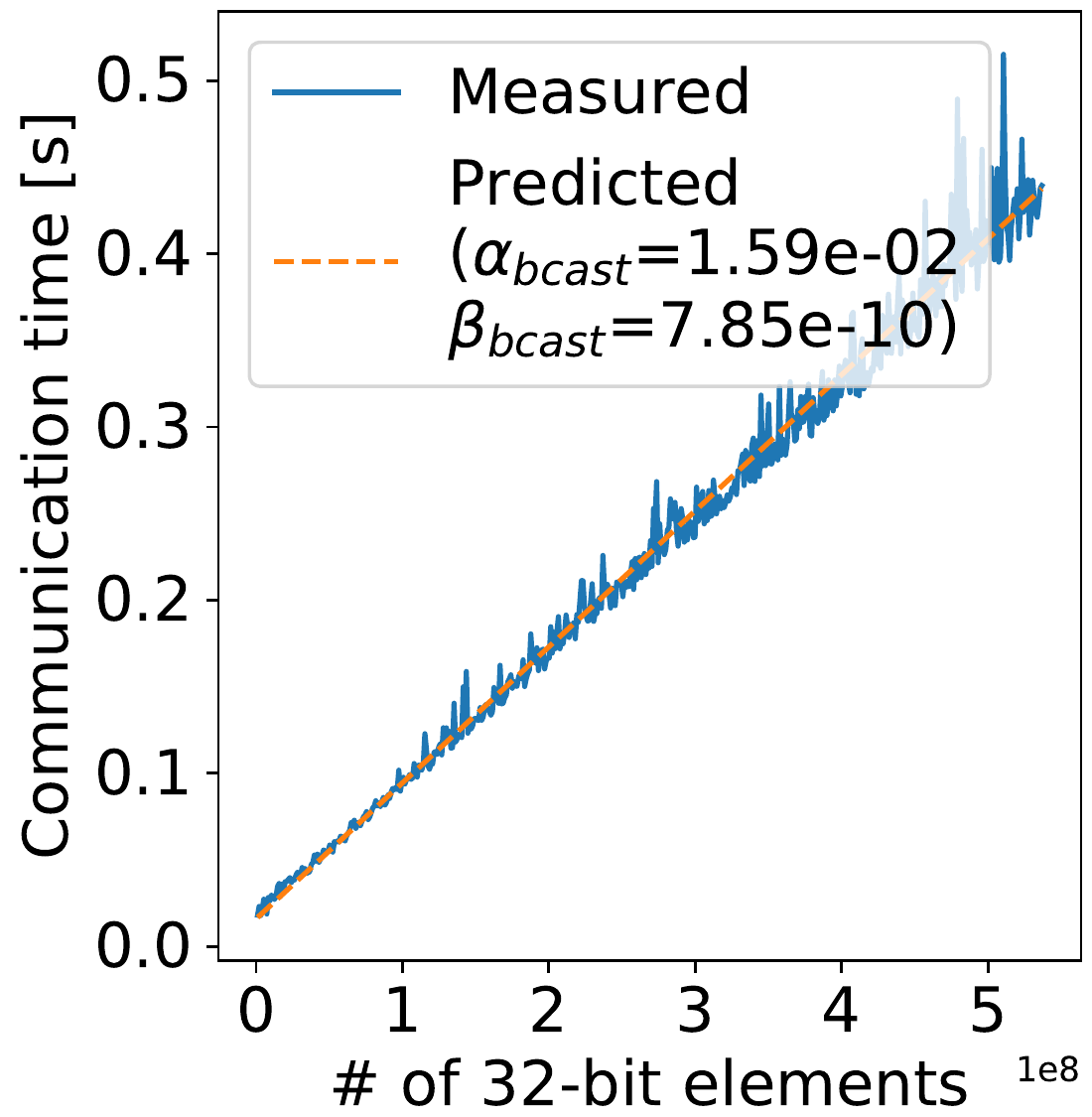}
		\caption{Broadcast.}
	\end{subfigure}
	
	\caption{The communication models of all-reduce and broadcast on our 64-GPU cluster. The communication time is the average of 100 runs with 10 times' warmup.}
	\label{fig:comm-models}
\end{figure}

\textbf{Communication models.} We measure the all-reduce and the broadcast operations on a range of message sizes in [1M, 512M] to estimate the parameters $\alpha_{ar}$ and $\beta_{ar}$ in Eq.~(\ref{equ:allreduce-model}) and $\alpha_{bcast}$ and $\beta_{bcast}$ in Eq.~(\ref{equ:broadcast-model}) using NCCL-2.4.7 and Horovod communication collectives. The results are shown in Fig.~\ref{fig:comm-models}, which indicates the parameters are well estimated.

\begin{figure}[!h]
	\centering
	\includegraphics[width=\linewidth]{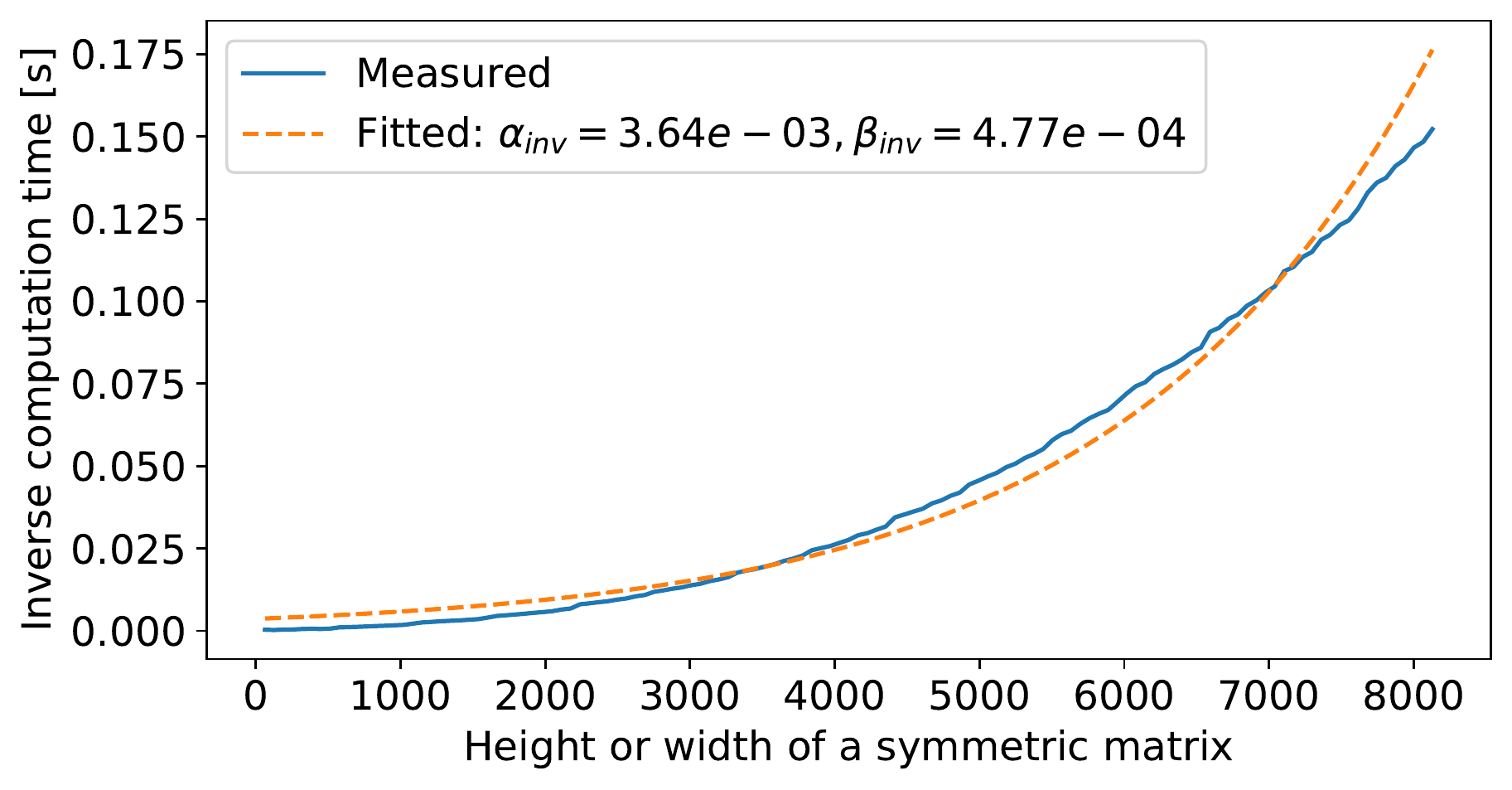}
	\caption{The computation time model of inverting a matrix on a GPU (Nvidia RTX2080Ti).}
	\label{fig:inv-model}
\end{figure}

\textbf{Computation model.} We benchmark the inverse operation on an RTX2080Ti GPU with a serial of symmetric matrices whose height or weight is in [64,8192] to estimate $\alpha_{inv}$ and $\beta_{inv}$ in Eq. (\ref{equ:inverting-model}). The computation time is also measured by the average time in 100 runs with 10 times' warmup. The result is shown in Fig.~\ref{fig:inv-model}, which indicates the exponential function of Eq. (\ref{equ:inverting-model}) well describes the computational cost of the inverse operation on a GPU.

\subsection{Wall-clock Iteration Time}
\begin{table}[!h]
    \centering
     \caption{Average iteration wall-clock time (in seconds) of 1,000 running iterations and speedups. $SP_1$ and $SP_2$ represent the speedup of SPD-KFAC over D-KFAC and MPD-KFAC, respectively.}
    \label{table:iteration-time}
    \centering
    \addtolength{\tabcolsep}{-0.5pt}
    \begin{tabular}{|c|c|c|c||c|c|}
    \hline
   Model & D-KFAC  & MPD-KFAC & SPD-KFAC & $SP_1$ & $SP_2$ \\\hline\hline
  ResNet-50 & 0.8525 & 0.7635 & 0.6755 & 1.26 & 1.13 \\\hline
  ResNet-152  & 1.5807 & 1.3933 & 1.1689 & 1.35 & 1.19 \\\hline
  DenseNet-201 & 1.4964 & 1.5340 & 1.3615 & 1.10 & 1.13 \\\hline
  Inception-v4  & 1.1857 & 1.1473 & 0.9907 & 1.20 & 1.16 \\\hline
    \end{tabular}
\end{table}
Since our SPD-KFAC can converge at the same speed in terms of the number of iterations as D-KFAC (or MPD-KFAC), we just need to compare the wall-clock iteration time with D-KFAC and MPD-KFAC on training DNNs. The results are shown in Table~\ref{table:iteration-time}, in which the numbers are the average of 1,000 iterations and 100 warmup iterations for each algorithm. The results show that our SPD-KFAC runs 10\%-35\% and 13\%-19\% faster than D-KFAC and the existing state-of-the-art solution MPD-KFAC, respectively.

\begin{figure}[!h]
	\centering
	\includegraphics[width=\linewidth]{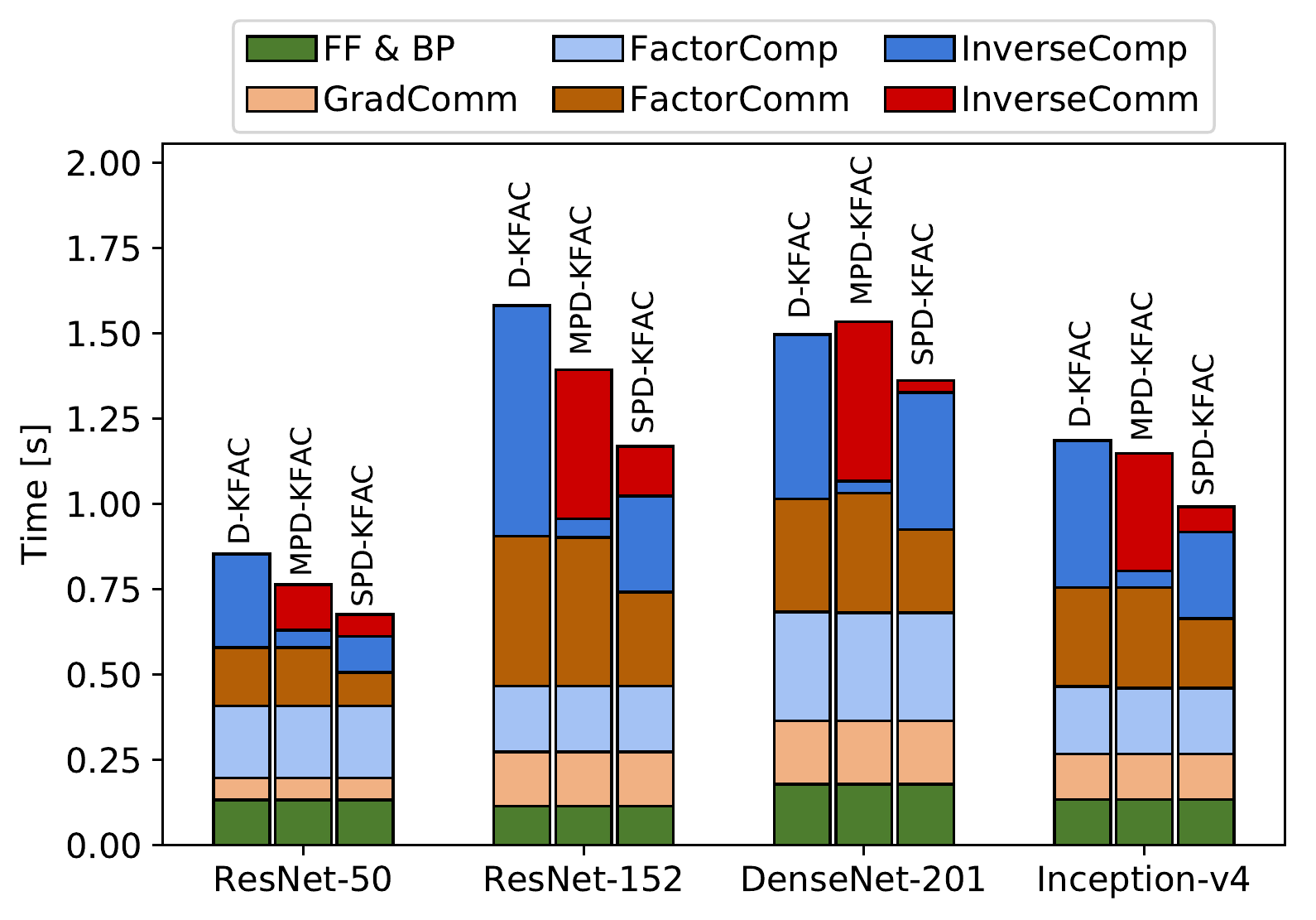}
	\caption{Time breakdowns of different algorithms.}
	\label{fig:spdfac-breakdowns}
\end{figure}

The time breakdowns are shown in Fig.~\ref{fig:spdfac-breakdowns}. Note that the feed-forward and backward computations (FF\&BP), gradient communications (GradComm), and Kronecker factors computations (FactorComp) are not optimized, so their time costs on different algorithms are the same. On one hand, in terms of the communication costs of aggregating Kronecker factors (FactorComm), we can see that our SPD-KFAC takes smaller time than MPD-KFAC because some communication overheads in SPD-KFAC can be hidden by the computations of Kronecker factors. We will dive into the details in the next subsection. On the other hand, regarding the benefits of the load-balancing placement, the computation time of inverting matrices (InverseComp) in SPD-KFAC is higher than MPD-KFAC, while SPD-KFAC has much smaller communication overheads (InverseComm) of broadcasting the inverted results to other GPUs than MPD-KFAC. In particular, MPD-KFAC is even slower than D-KFAC on DenseNet-201 due to the high communication overheads of broadcasting inverses, while our SPD-KFAC can avoid such a sacrifice by the load-balancing placement and hence SPD-KFAC achieves improvement over D-KFAC.

In summary, both the pipelining and load-balancing placement algorithms contribute to the performance improvement of our SPD-KFAC.

\subsection{Benefits of Pipelining}
\begin{figure}[!h]
	\centering
	\includegraphics[width=\linewidth]{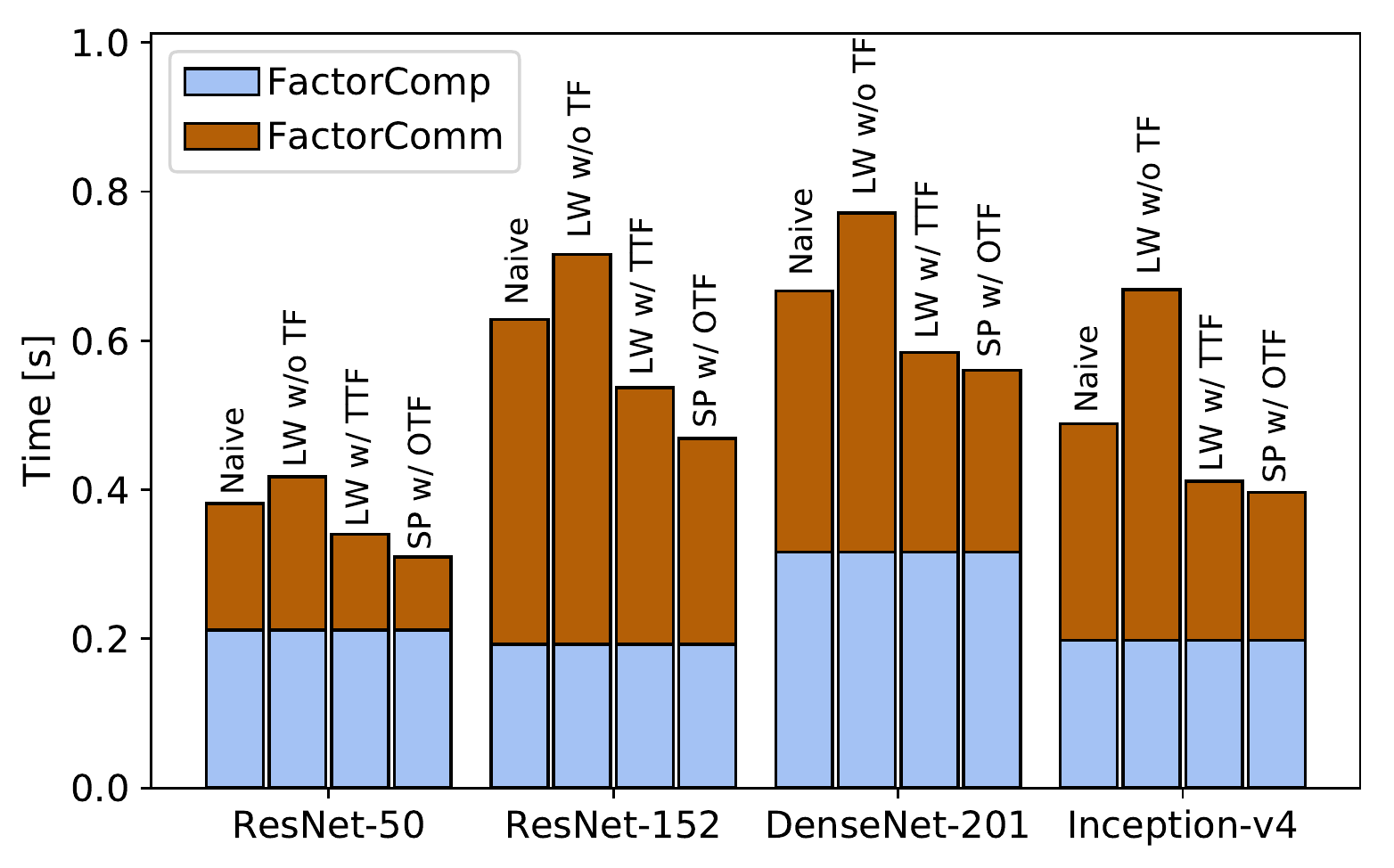}
	\caption{Comparison of pipelining between computations and communications of Kronecker factors.}
	\label{fig:pipelining-results}
\end{figure}

We dive deep into understanding the performance benefits of pipelining computation and communication of Kronecker factors. We measure the Kronecker factor computation time and non-overlapped communication time during training for each algorithm. Note that the non-overlapped communication time is the elapsed time of communication whose overlapped parts are excluded. To show the effectiveness of our smart parallel with optimal tensor fusion (SP w/ OTF), we compare with the naive version (Naive) of pipelining between the communications of $A$ and the computations of $G$, the layer-wise pipelining method without tensor fusion (LW w/o TF), and the layer-wise pipelining method with tensor fusion by a threshold\footnote{Default 64MB in Horovod} (LW w/ TTF). The results are shown in Fig.~\ref{fig:pipelining-results}. It can be seen that LW w/o TF is worse than Naive, which indicates that the layer-wise communication without tensor fusion would result in larger overheads due to the startup time of all-reduce operations. Using the default tensor fusion with a pre-defined threshold (LW w/ TTF) would achieve better efficiency than Naive, but it is still sub-optimal. Our algorithm with the optimal tensor fusion solution can achieve further improvement. As a result, our pipelining method can hide more 50\%-84\% communication overheads of aggregating Kronecker factors in the four deep models than the overlapping solution from~\cite{pauloski2020convolutional, ueno2020rich}.

\subsection{Benefits of Load-Balancing Placement}
\begin{figure}[!h]
	\centering
	\includegraphics[width=\linewidth]{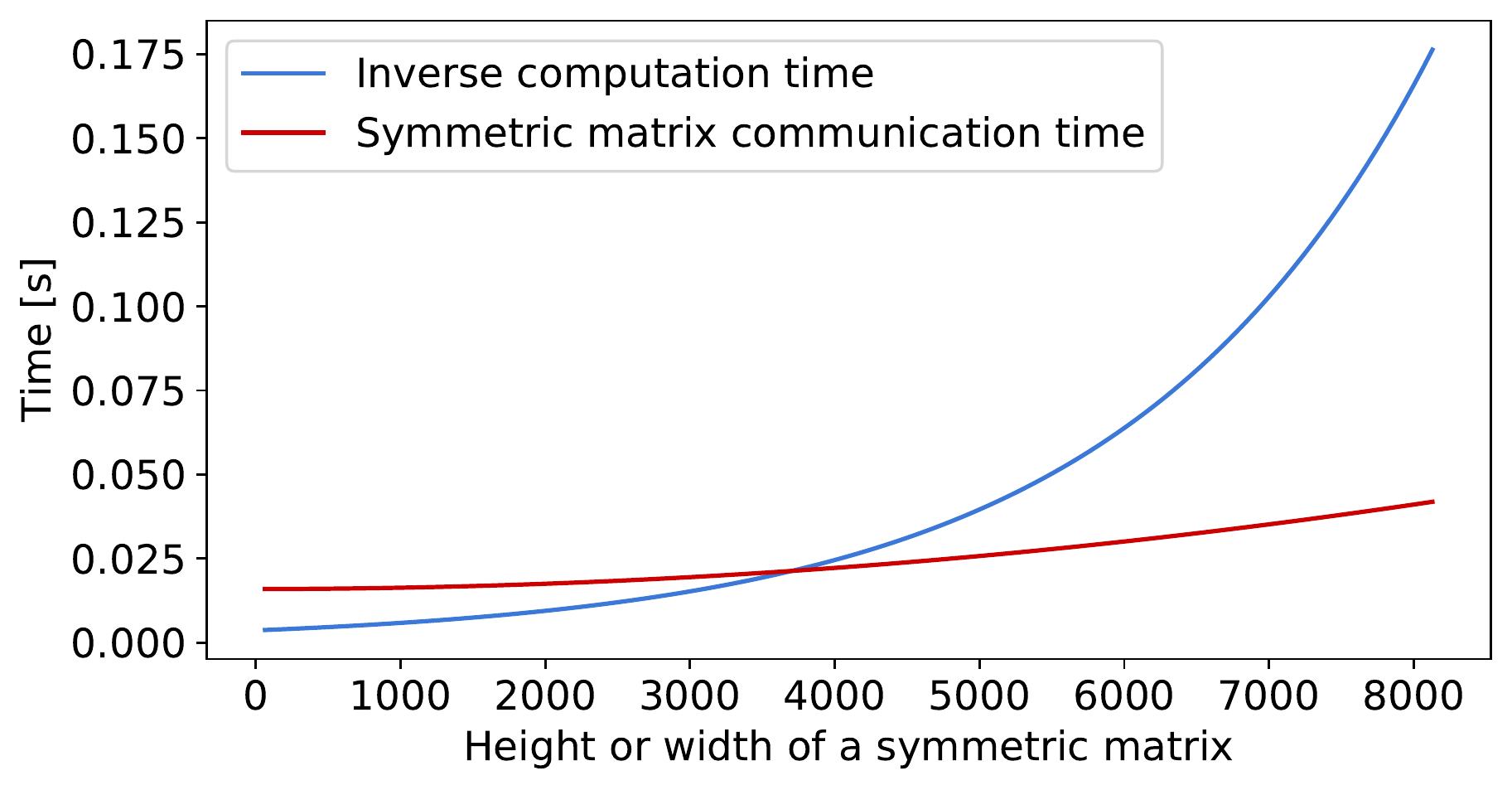}
	\caption{Comparison of computation and communication models on our 64-GPU (RTX2080Ti) cluster.}
	\label{fig:inv-vs-bcast}
\end{figure}

Our proposed load-balancing placement with considering the communication cost benefits from load-balancing placement and determining small tensors to be inverted on all GPUs. According to the computation model of Eq. (\ref{equ:inverting-model}) and the communication model of Eq. (\ref{equ:broadcast-model}) with the estimated parameters in Fig.~\ref{fig:inv-model} and Fig.~\ref{fig:comm-models}(b) respectively, we compare these two models in Fig.~\ref{fig:inv-vs-bcast}. We can see that when the tensor dimension is smaller than a threshold, it would be better to make the tensor be NCT. 

\begin{figure}[!h]
	\centering
	\includegraphics[width=\linewidth]{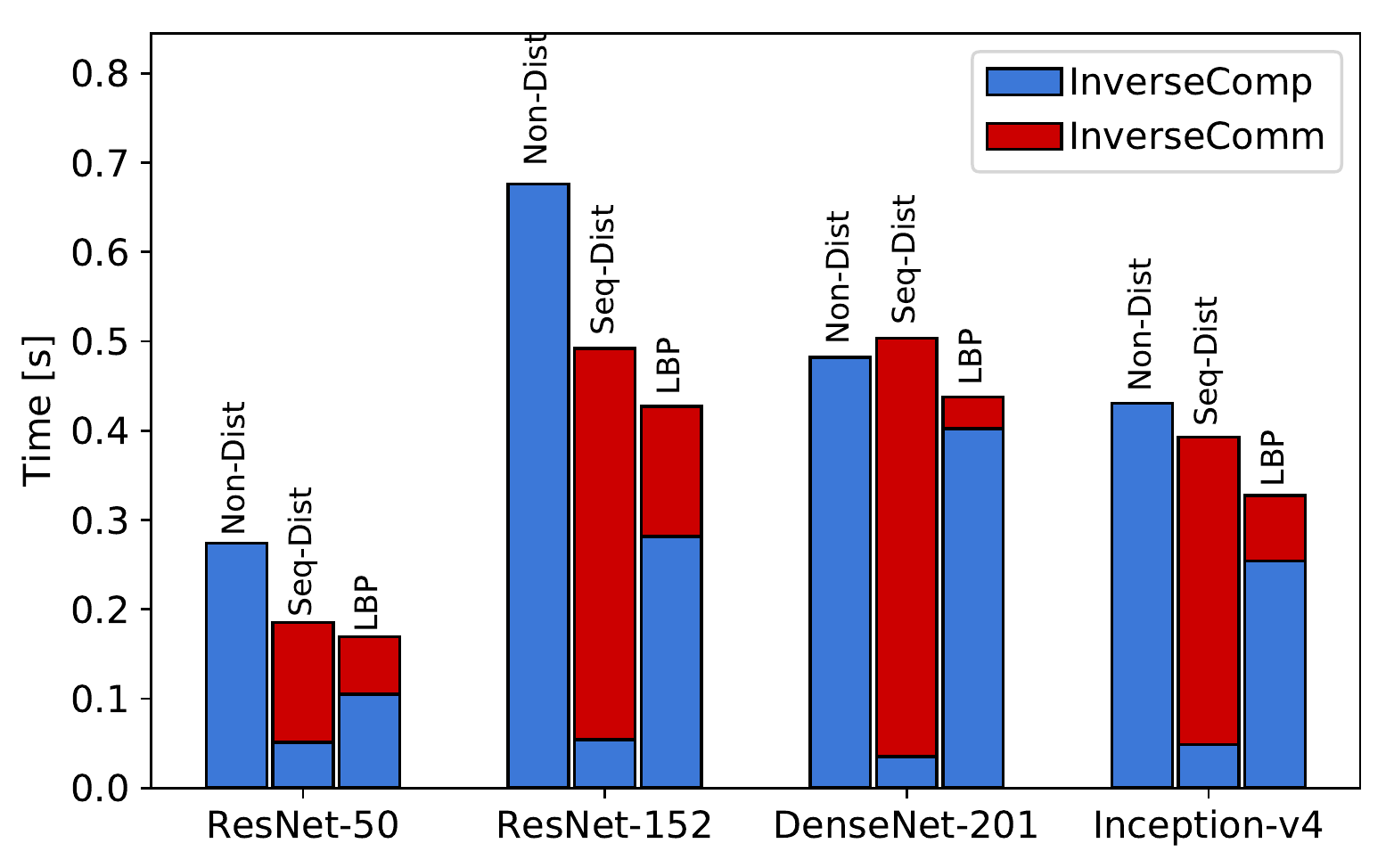}
	\caption{Comparison of inverting multiple matrices.}
	\label{fig:bwp-results}
\end{figure}

Regarding our proposed load-balancing placement (LBP), we compare the wall-clock time of inverting Kronecker factors with the naive version without distributing the workloads (Non-Dist) to multiple GPUs and the distributed version that places the workloads sequentially (Seq-Dist)~\cite{osawa2019large,pauloski2020convolutional,ueno2020rich}. The results are shown in Fig.~\ref{fig:bwp-results}, which indicates that our LBP is always better than Non-Dist and Seq-Dist. Seq-Dist alleviates the computation overheads of inverting matrices on each GPU, but it brings significant communication costs. In DenseNet-201, Seq-Dist is even worse than Non-Dist. Our LBP balances the communication and computation costs to place the workloads on multiple GPUs such that the overall time can be reduced. As a result, our LBP achieves 10\%-62\% improvement in computing the inverses of Kronecker factors over existing Non-Dist and Seq-Dist solutions.

\subsection{Ablation Results}
\begin{figure}[!h]
	\centering
	\includegraphics[width=\linewidth]{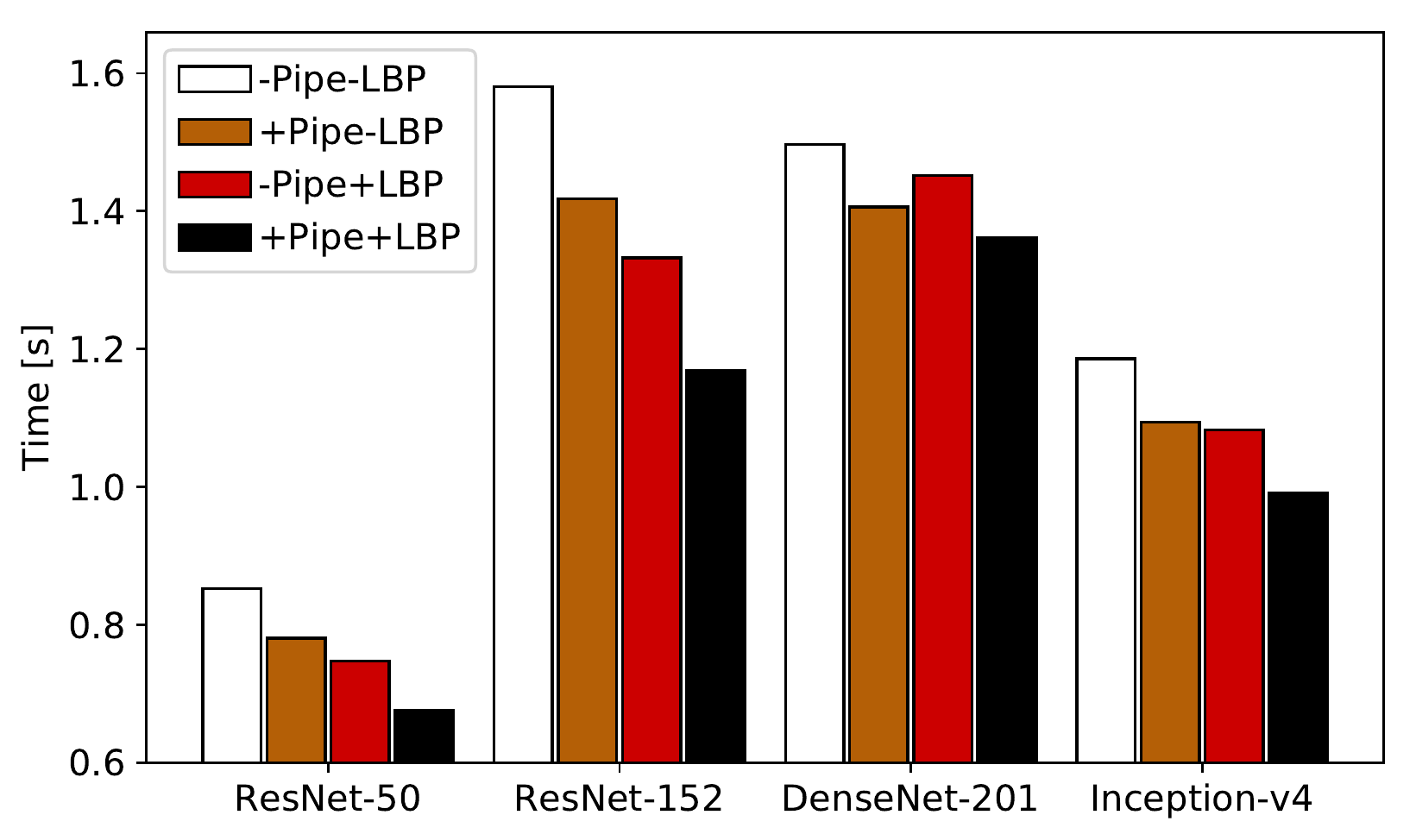}
	\caption{The ablation results with particular optimizations.}
	\label{fig:step-by-step-results}
\end{figure}

\begin{table}[!ht]
	\centering
		\caption{Notations with particular optimizations}
		\label{table:notations}
    \begin{tabular}{|c|c|c|}
    	\hline
        Name & w/ Pipelining (\S\ref{subsec:pipelining}) & w/ LBP (\S\ref{subsec:lbp}) \\\hline\hline
        -Pipe-LBP & \ding{55} & \ding{55} \\\hline
        +Pipe-LBP & \ding{51} & \ding{55} \\\hline
        -Pipe+LBP & \ding{55} & \ding{51} \\\hline
        +Pipe+LBP & \ding{51} & \ding{51} \\\hline
    \end{tabular}
\end{table} 
We provide the ablation results of our proposed SPD-KFAC. For ease of presentation, some notations to represent particular optimizations are shown in Table~\ref{table:notations}, and the time performance is shown in Fig.~\ref{fig:step-by-step-results}. With only pipelining, +Pipe-LBP achieves about 10\% improvement, and with only LBP, -Pipe+LBP achieves 3\%-18\% improvement. Putting the two optimizations together, +Pipe+LBP achieves 10\%-35\% improvement. 

\section{Conclusion}\label{sec:conclusion}
In this paper, we first identified the performance bottlenecks of the distributed second-order learning algorithm, D-KFAC, on training DNNs. We then introduced our D-KFAC algorithm named SPD-KFAC with smart parallelism between computing tasks and communication tasks by proposing two novel optimization techniques to improve the training efficiency on GPU clusters: 1) making Kronecker factors' communications be pipelined with their computations to hide some communication overheads, where the optimal tensor fusion technique is applied to make the best overlapping between communications and computations, and 2) making the multiple tensors computations (inverting matrices) on a GPU cluster be load-balancing while incorporating the communication costs. We conducted extensive experiments on a 64-GPU cluster connected with 100Gb/s InfiniBand with four popular DNNs. The experimental results showed that our SPD-KFAC outperforms the existing state-of-the-art solutions.
\section*{Acknowledgments}
The research was supported in part by RGC RIF grant R6021-20, and RGC GRF grants under the contracts 16207818 and 16209120. The research was also in part sponsored by CCF-Baidu Open Fund (No. CCF-BAIDU OF2020017).


\bibliographystyle{IEEEtran}
\bibliography{cites}

\end{document}